\documentclass[runningheads]{llncs}
\usepackage[T1]{fontenc}
\usepackage{tabularx}
\usepackage{graphicx}
\usepackage{hyperref}
\usepackage{color}

\usepackage{verbatim} 
\begin{document}

\title{Unravelling Organisational Rule Systems in Requirements Engineering}

\titlerunning{Unravelling Organisational Rule Systems}  
%
%
\author{Jöran Lindeberg\inst{1}\orcidID{0000-0001-7806-749X} \and
Eric-Oluf Svee\inst{1}\orcidID{0000-0003-2218-8094} \and
Martin Henkel\inst{1}\orcidID{0000-0003-3290-2597}} 
\authorrunning{Lindeberg et al.}
%
\institute{Department of Computer and Systems Sciences (DSV), Stockholm University, Borgarfjordsgatan 12, Kista, Stockholm
 } 
\maketitle              

\begin{abstract}

\emph{Context and motivation}: Requirements engineering of complex IT systems needs to manage the many, and often vague and conflicting, organisational rules that exist in the context of a modern enterprise. At the same time, IT systems affect the organisation, essentially setting new rules on how the organisation should work.  

\emph{Question/problem}: Gathering requirements for an IT system involves understanding the complex rules that govern an organisation. The research question is: How can the holistic properties of organisational rules be conceptualised? 

\emph{Principal ideas/results}: This paper introduces the concept of organisational rule systems that may be used to describe complex organisational rules. The concept and its components are presented as a conceptual framework, which in turn is condensed into a conceptual framework diagram. The framework is grounded in a critical literature review. 

\emph{Contribution}: The conceptual framework will, as a first step of a wider research agenda, help requirements engineers understand the influence of organisational rules.

\end{abstract}
\section{Introduction} \label{introduction}

As recognised by Jarke et al. \cite{jarke2011brave}, complex IT systems and their organisational context are \emph{intertwined}. In their words, "the implementation impacts not only the technical system, but also their organizational and social settings and often increases interaction complexity in unpredictable ways” \cite[p. 1000]{jarke2011brave}. In a similar vein, Baskerville et al. \cite[p. 28]{baskerville_digital_2019} use the concept of \emph{ontological reversal} to explain that "increasingly in our digital world, a digital version of reality is created first, and the physical version second (if needed)". One of their conclusions is that information systems (IS) scholars should take more holistic perspectives and increasingly view IT systems as part of socio-technical systems. Accordingly, the focus of requirements engineering (RE) has shifted from individual IT systems towards software-intensive ecosystems consisting of not only IT, but also e.g. social agents, organisational rules and regulatory issues. Hence, requirements engineers must not only understand organisational rules in order to implement them but also be aware of how the IT systems they design generate new rules that feedback into the organisational context. 

One aspect of the organisational context are the ubiquitous rules that govern modern enterprises. According to Semantics of Business Vocabulary And Business Rules (SBVR), a rule is “an element of guidance that introduces an obligation or a necessity.” \cite[Annex E, p. 4]{noauthor_semantics_2019}. Rules are effective because, unlike direct commands, they scale well across space, people, and time. Rules can also reduce the complexity that an organisational unit has to manage; Rules help to filter irrelevant information, and increase the power of decision-makers who, thanks to rules and routines, can control complex operations. In summary, rules can help create order from disorder \cite{ross_rules_2023} and are a form of exercising control. 

The effectiveness of rules has contributed to their popularity. Thus, identifying all relevant regulations for a large IT project can be far from trivial, not to mention discovering the requirements therein \cite{otto2007addressing}. Furthermore, while some rules are straightforward to implement, others are vague or conflicting, also leading to conflicting requirements. For example, in an ongoing case study \cite{fast_lappalainen_digitalisering_2021} of the healthcare system in Region Stockholm, Sweden, we have observed that the rules that regulate the use of IT in healthcare are relatively opaque, multidimensional, and of high complexity. The actually available legal space in healthcare is further reduced by both organisational and technical challenges.  

Many aspects of organisational rules have been researched in the field of business rules management. However, business rules management is closely associated with decision management in tightly specified situations \cite{zoet2014methods}, but much less so with broader, more comprehensive perspectives.

The research question of this paper is: How can the holistic properties of organisational rules be conceptualised? To that end, we introduce the concept of \emph{organisational rule system} and explicate its components in the form of a conceptual framework. The wider aim is to initiate a research agenda for organisational rule systems.

The remainder of this paper is structured as follows: Section \ref{methodolodgy} explains the methodology of a literature review, Section \ref{results} presents the results in the form of a conceptual framework, Section \ref{discussion} discusses the implications for RE, and concludes. 

\section{Methodology} \label{methodolodgy}

As explained by Grant \& Booth \cite{grant2009typology}, a critical literature review is typically not as systematic as other forms of reviews and may lack explicit criteria for search and analysis. Rather, the focus is on conceptual innovation based on existing, sometimes competing, schools of thought \cite{grant2009typology}. In the present work, the reviewed literature gravitated around the intersection of business rules management, Management and Organisation Studies (MOS), systems thinking, history, and design science. Numerous queries were tested using Google Scholar to identify literature within or with relevance for this area. Many sources were subject to backward and forward snowballing of citations. The result was a conceptual framework, constructed using three parts from different sources:
\begin{enumerate}
    \item  \textit{Individual Rules}: Definition of the core concept of rules. This was largely drawn from SBVR and its sister standard Business Motivation Model (BMM).
    \item  \textit{Properties of Rules}: Described facets of the rules, such as their level of enforcement. SBVR and BMM do not contain a full set of rule properties: therefore, properties from other sources were added, e.g. durability and rationale. 
    \item \textit{Rules as a Part of Systems}. This part of the framework describes how rules interact with other rules, and also the organisational context mentioned in Section \ref{introduction}. Sources for this framework part were works about socio-technical systems and how rules affect an organisation.
\end{enumerate}

\section{Organisational Rule Systems: A Conceptual Framework} \label{results}

\begin{figure}
\centerline{
\includegraphics[width=1.4\textwidth]{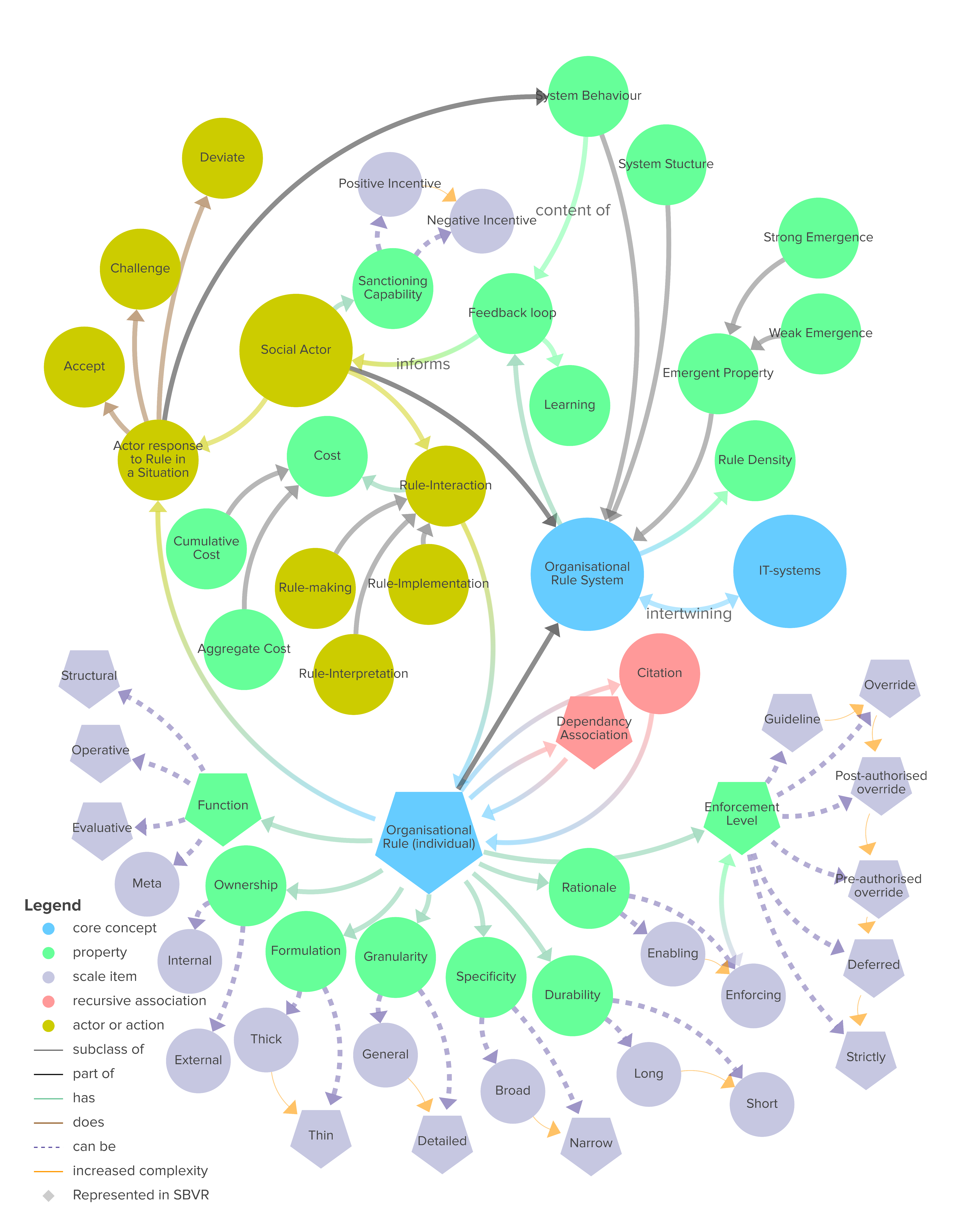}}
\caption{Diagram of the conceptual framework of organisational rule systems. It must be read together with the main text to be understandable. Note that concepts represented in, or associated with, SBVR are pentagons, while all other concepts are circles.} \label{fig:fig-framework}
\end{figure}

This section presents the conceptual framework, as shown in Figure \ref{fig:fig-framework}, and also online \cite{lindeberg_conceptual_2023}. We start with individual rules and their properties as a foundation for system perspectives. Each concept that is part of the framework is marked with \textit{italics}.

\subsection{Individual Rules}

SBVR offers a structured and well-known way to describe rules, which makes this influential business rules management standard a suitable point of departure of the conceptual framework. 

What distinguishes a rule from other forms of guidance (such as values and goals) is that it sets boundaries of the action space of an actor in a certain situation. Since our domain of interest is organisations, we use the term \emph{ organisational rule} (equivalent to an SBVR directive). An organisational rule can be either a policy or a business rule. Organisational rules that are not practicable are policies. Being practicable means that the intended interpreters can know with some certainty what to do when applying the rule in practice. Many, but not all, rules that are practicable can be automated, that is, interpreted by machines. 

\subsection{Properties of Rules}

Most of the properties presented here will be expressed as dichotomies, even though, on a closer look, they are rather opposites on a scale. Note that some properties are from SBVR, and some from other sources.  

Rules can be classified according to their function: A \emph{structural rule} provides definitions of reality, e.g. which persons an enterprise considers as its costumers. A definition in a business vocabulary is a type of structural rule; An \emph{operative rule} regulates what is allowed; An \emph{evaluative rule} \cite{Burns1987-vv} is close to what BMM denominates as goal. In other words, the line between rule and goal is elusive; A \emph{meta-rule} is a rule about rules. An example of meta-rules, observed by most people, is that in general it is best to follow rules rather than breaking or trying to change them \cite{Burns1987-vv}.

In particular in highly regulated sectors, such as healthcare, the purpose of many \emph{internal rules} is to comply with \emph{external rules}. While an organisation has \emph{ownership} over its internal rules, the opposite is true for external rules. A common type of external rule is regulation. Nevertheless, there are also external rules that are not regulations, but are still more mandatory than not, such as influential standards \cite{zoet2014methods}.

Rules can be thin or thick in their \emph{formulation}. According to Daston, a \emph{thick rule} "is upholstered with examples, caveats, observations, and exceptions" \cite[p. 3]{daston_rules_2022}. It presumes a messy world for which the rules must be adapted to the circumstances. In contrast, \emph{thin rules} "implicitly assume a predictable, stable world in which all possibilities can be foreseen" and "do not invite the exercise of discretion" \cite[p. 3]{daston_rules_2022}. The thinnest of rules would be e.g. computer code. For thin rules to work, materials and measures must be standardised. \cite{daston_rules_2022}. 

Rules can have more or less \emph{granularity}, i.e. level of detail \cite{daston_rules_2022}. Rules can cover very wide domains, e.g. universal human rights law, or have very high \emph{specificity}, such as the permitted content of a particular form field. Rules with wide domains tend to be less granular \cite{daston_rules_2022}. The \emph{durability} of a rule expresses how long it remains unchanged \cite{zhu_dynamics_2019}.

In an organisational context, rules are a type of formalisation. Formalisation is "the extent of written rules, procedures, and instructions" that guide organisational processes \cite[p. 62]{adler_two_1996}. 

The type of formalisation---a rule's \emph{rationale}---can be \emph{enabling} or \emph{coercive}. Adler \& Borys have compared formalisation with the introduction of new technology \cite{adler_two_1996}. They argue that coercive technology is designed "with a fool-proofing and deskilling rationale, aiming to reduce reliance on more highly paid, highly skilled, and powerful workers" \cite[p. 67]{adler_two_1996}. In contrast, enabling technology "can be designed with a usability and upgrading rationale, aiming to enhance users' capabilities and to leverage their skills and intelligence" \cite[p. 67]{adler_two_1996}. The same distinction can be made with regard to formalisation, Adler \& Borys argue. 

The saying that "rules are rules" (and should be obeyed at all times) is a simplification. As a matter of fact, rules have different enforcement levels both in theory and in practice \cite{ross_rules_2023}. To express this, SBVR has borrowed an ordinal scale from BMM, ranging from a mere \emph{guideline} without enforcement, different forms of overrides (breake the rule but explain why), \emph{deferred enforcement}, and finally \emph{strict enforcement}. This last alternative can also be called rigid application, while the opposite side of the scale leaves more room for discretion \cite{daston_rules_2022}. 

Drawing from Daston \cite{daston_rules_2022}, it can be noted that rules that are strictly enforced with negative incentives, while also being thin, detailed and narrow, can be very efficient. On the other hand, they leave less room for human judgement, lead to \emph{increased complexity}, and must be most carefully designed to work as expected. Moreover, even then such rule systems do manage to foresee all situations that may arise, they must be changed as soon as the environment changes, shortening their durability. This constitutes a challenge since, as observed by Jarke et al. \cite{jarke2011brave}, the rate of change in the organisational context must be matched by the designers of its IT systems.

\subsection{Rules as Part of Systems}

The interconnections between rules, and between rules and actors, form an \emph{organisational rule system} that as stated is Section \ref{introduction} is intertwined with \emph{IT systems}.  

Rules can be interconnected with other rules through \emph{citations} \cite{zhu_dynamics_2019}. A rule that affects the meaning of another rule is a \emph{dependency} rule in SBVR. 

Detailed rules are associated with \emph{rule density}, i.e. to what extent a certain domain is controlled by rules. Put differently, a heavily regulated industry is a domain with a high rule density.

Rules interact with \emph{social actors} such as humans \cite{zoet2014methods}. In addition to being controlled by rules, they can do \emph{rule-making}, \emph{rule-interpretation}, and \emph{rule-implementation} \cite{Burns1987-vv}. When faced with a rule in a given situation, a social actor can choose between \emph{accept}, \emph{deviate}, or \emph{challenge} the rule \cite[p. 53]{Burns1987-vv}. Actors can also have a \emph{sanctioning capability} \cite[p. 79]{Burns1987-vv}, including both positive and negative \emph{incentives} \cite[p. 7]{ross_rules_2023}. 

One reason for not accepting rules is that they have \emph{costs} for actors. Costs can be both \emph{aggregate} and \emph{cumulative}. As explained by Sunstein \cite[p. 588]{sunstein_regulatory_2014} aggregation means to simply sum up the costs of the different rules that apply at a specific period of time. In contrast, a cumulative perspective of burdens in this context takes into account that each receiving actor (e.g. employee or organisational unit) only has a certain amount of resources (e.g. number of work hours per week) for its disposal, and each added rule will decrease time spent on other opportunities. 

Like any other system, some properties of an organisational rule system will be \emph{emergent}, different from the sum of its parts. There are two forms of emergence: weak and strong \cite{colchester_systems_2016}. \emph{Weak emergence} derives from the relationships between the components, and can, in theory, be computed if all information about the system is available. \emph{Strong emergence} goes beyond this and cannot be predicted: it can only be detected after the fact through \emph{feedback} (i.e. impact assessment \cite{zoet2014methods}), which is necessary for \emph{learning} \cite{argyris_double_1977}, either of the more shallow \emph{single-loop} type or the more profound \emph{double-loop}. Feedback and learning are of great importance, because, as pointed out by Jarke et al. \cite{jarke2011brave}, the behaviour of complex organisations is unpredictable. 

 It can be worth noting that from the perspective of systems thinking \cite{colchester_systems_2016} and its subfield organisational cybernetics \cite{rios_systems_2012} it is hardly surprising that many organisational rule systems are malfunctioning. A system is composed of a hierarchy of sub-systems, which need an amount of self-organisation. Many rules, necessary or not, will impede the self-organisation of the subsystems. Since self-organisation is fundamental for resilience and long-term survival, there are reasons to be cautious before tackling a problem with more rules. In fact, self-organisation requires a certain amount of disorder \cite{meadows_thinking_2008}. Also, in a nearly empty world, with loosely coupled systems, it is easier to do local changes without negative effects \cite{march_organizations_1993}. In other words, the more rules, the more risk of unexpected behaviour. Furthermore, today's complex organisations are rife with inherent paradox \cite{berti_dark_2021}. It is also worth remembering that the bounded rationality \cite{simon_sciences_1996} of the human mind limits our ability to make well-founded choices, even when all information is available. 

\section{Discussion and Conclusion: Rules and Requirements} \label{discussion}

As can be observed in the conceptual framework, there are considerable aspects of organisational rules systems that are not supported by current business rule management. SBVR lacks the systems thinking perspective that we argue must be addressed. Differently put, it implicitly assumes a deterministic and rational world, making it inadequate for addressing the holistic and dynamic perspectives identified by Jarke et al. \cite{jarke2011brave} in Section \ref{introduction}. 

To stay relevant, requirements engineering will need to increasingly embrace these perspectives and be able to switch lenses according to the problem at hand. However, it should also be conceded that recognising the intertwining between rules in IT systems and rules in the organisational environment also has risks. Perspectives that seek to represent reality in all its complexity will themselves become too complex to be useful. There is a balance to be struck here.  

The extent of the identified gap calls for a research agenda. As a first step towards that end, the conceptual framework in this paper constitutes a preliminary \emph{theory of the problem} \cite{majchrzak_designing_2016}. A second step would be to better understand the state-of-the-art of research fields adjacent to organisational rule systems. In this paper, SBVR was used as a representative of business rules management, but a systematic literature review of existing approaches is needed. A third step would be to better ground the theory of the problem in practice. This aim could be done through case studies and surveys of how organisational rule systems affect the requirements engineering of particular IT systems in highly regulated industries. A fourth step would switch the focus to a \emph{theory of a solution} \cite{majchrzak_designing_2016} by proposing formal modelling methods that make complex organisational rule systems more comprehensible, thereby supporting requirements engineering of new IT systems. A fifth step would be evaluating and comparing said methods, e.g. through case studies.  
 
To conclude, this paper presented an initial conceptual framework describing rules, based on a critical literature review, and argued that organisational rule systems and their implications for requirements engineering deserve more attention. A limitation of this study is its lack of systematic criteria for selection and analysis. A future systematic literature review would be a useful complement, and further contribute to the proposed research agenda. 

%
%
%
 \bibliographystyle{splncs04}
%
\bibliography{references2}
\end{document}